\documentclass[aps,twocolumn,epsf,showpacs]{revtex4}
\usepackage{graphicx}

\begin{document}

\title{Miscibility in a degenerate fermionic mixture induced by linear
coupling}
\author{Sadhan K. Adhikari}
\affiliation{Instituto de F\'{\i}sica Te\'{o}rica, UNESP -- S\~{a}o Paulo State
University, 01.405-900 S\~{a}o Paulo, S\~{a}o Paulo, Brazil }
\author{Boris A. Malomed}
\affiliation{Department of Interdisciplinary Studies, School of Electrical Engineering,\\
Faculty of Engineering, Tel Aviv University, Tel Aviv 69978, Israel}

\begin{abstract}
We consider a one-dimensional mean-field-hydrodynamic model of a
two-component degenerate Fermi gas in an external trap, each component
representing a spin state of the same atom. We demonstrate that the
interconversion between them (linear coupling), imposed by a resonant
electromagnetic wave, transforms the immiscible binary gas into a miscible
state, if the coupling constant, $\kappa $, exceeds a critical value, $%
\kappa _{\mathrm{cr}}$. The effect is predicted in a variational
approximation, and confirmed by numerical solutions. Unlike the recently
studied model of a binary BEC with the linear coupling, the components in
the immiscible phase of the binary fermion mixture never fill two separated
domains with a wall between them, but rather form anti-locked ($\pi $%
-phase-shifted) density waves. Another difference from the bosonic mixture
is spontaneous breaking of symmetry between the two components in terms of
numbers of atoms in them, $N_{1}$ and $N_{2}$. The latter effect is
characterized by the parameter $\nu \equiv (N_{1}-N_{2})/\left(
N_{1}+N_{2}\right) $ (only $N_{1}+N_{2}$ is a conserved quantity), the onset
of miscibility at $\kappa \geq \kappa _{\mathrm{cr}}$ meaning a transition
to $\nu \equiv 0$. At $\kappa <\kappa _{\mathrm{cr}}$, $\nu $ features
damped oscillations as a function of $\kappa $. We also briefly consider an
asymmetric model, with a chemical-potential difference between the two
components. The relation between the imbalance in the spin population,
induced by the linear coupling, and the developing spatial structure
resembles the known Larkin-Ovchinnikov-Fulde-Ferrell (LOFF) states in the
Fermi mixture. Dynamical states, when $\kappa $ is suddenly switched from$\ $%
zero to a value exceeding $\kappa _{\mathrm{cr}}$, are considered too. In
the latter case, the system features oscillatory relaxation to the mixed
state.
\end{abstract}

\maketitle

\section{Introduction}

Experimental and theoretical studies of degenerate Bose \cite{books} and
Fermi \cite{Fermi} gases at ultra-low temperatures have become a vast
research area. This field offers a unique testbed for the implementation of
various ideas from classical and quantum nonlinear dynamics. In particular,
experiments are possible in binary mixtures of atoms in two different
hyperfine (spin) states. The mixtures were created by means of the
sympathetic-cooling technique in the Bose-Einstein condensate (BEC) made of $%
^{87}$Rb atoms \cite{RbSympathetic}. In such a binary condensate, mutual
interconversion, i.e., linear coupling between the two species, can be
induced by means of a resonant (spin-flipping) electromagnetic wave, with a
frequency in the GHz range \cite{NZ}. Various effects were considered in
this setting, including Josephson oscillations between the two states \cite%
{Josephson}, domain walls between them \cite{DomainWalls},
``breathing-together" oscillatory modes in the mixture \cite{Kennedy}, and
nontopological vortices \cite{latest}.

In a recent work \cite{Ilya}, a new effect was predicted, namely, transition
to miscibility induced by the linear coupling between the immiscible
components in a binary condensate. It was demonstrated that, essentially, it
occurs as a transition from a state with the two components separated by a
domain wall to a completely mixed one. Quantitatively, the transition is
characterized by an order parameter which determines the separation between
centers of mass of the two components. With the increase of the
linear-coupling strength (coupling constant), the order parameter decreases,
and the full mixing sets in when it vanishes. Experimental observation of
this effect is feasible due to the fact that the mixture of two hyperfine
states in $^{87}$Rb is immiscible, but close to the miscibility threshold
\cite{RbSympathetic,proximity}. The proximity of the mixture to the
threshold can be additionally controlled by means of the Feshbach resonance,
which can be induced by an external magnetic field \cite{Feshbach} or
optical pump \cite{OpticalFeshbach}.

Phase separation (demixing) due to the repulsion between two components was
also investigated in boson-fermion mixtures \cite{BFmixture}, and, more
recently, in a mixture of two fermion species \cite{static,Sadhan1}. In the
latter case, a situation is possible when the species are two different spin
states of the same fermion atom (as realized experimentally in $^{40}$K \cite%
{K} and $^{6}$Li \cite{Li}), hence the linear interconversion between them
can be induced by the properly tuned electromagnetic wave.

In a recent experimental work \cite{Stoof}, the population
imbalance between two spin states (``polarization") in a
degenerate $^{6}$Li gas (trapped in a cigar-shaped cavity) was
controlled by means of this technique. It was observed that, due
to strong attraction between the different states (induced by the
Feshbach resonance in external magnetic field) they formed a fully
mixed state in the central part of the trap, that was separated by
two sharp boundaries from side domains, filled with the atoms
representing a majority state (see Figs. 3 and 5 in Ref.
\cite{Stoof}).

Thus, it is relevant to consider the transition to mixing in the binary
fermionic gas under the action of the linear coupling (as well as in the
bosonic mixture, the proximity to the miscibility point may be additionally
controlled by means of the Feshbach resonance, which affects the strength of
the repulsion between the two species \cite{fsff}). This is the subject of
the present work.

Rigorous description of the dynamics of the degenerate gas of fermions is
based on the system of quantum-mechanical equations of motion for individual
fermions (see, e.g., Ref. \cite{Mario}), which is difficult to handle if the
gas contains many atoms. Simplified approaches, of the
mean-field-hydrodynamic (MFHD) type, were also elaborated for the
description of quantum fermionic gases, in static \cite{static} and
dynamical \cite{hydro,Sadhan-collapse,Sadhan-solitons,Sadhan1} settings. In
the latter case, the approximation amounts to a single evolution equation
for the fermion wave function, which formally resembles the Gross-Pitaevskii
equation (GPE) for the boson gas \cite{books}, but features a different
nonlinearity. This MFHD equation was successfully used to describe collapse
\cite{Sadhan-collapse} and solitons \cite{Sadhan-solitons} in degenerate
fermionic gases, and a system of nonlinearly coupled equations was employed
to predict the phase separation in the binary gas with repulsion between the
species \cite{Sadhan1}.

In this work, we aim to investigate the onset of mixing in the immiscible
binary fermion gas under the action of the extra linear coupling, in the
(effectively) one-dimensional situation. In Sec. II, we formulate the model,
and then elaborate an analytical approach to the problem, based on a
variational approximation (VA). The latter provides a description of the
binary mixture in terms of the number-of-atoms ratio in the two components, $%
N_{1}/N_{2}$. In the presence of the linear mixing, only the total number is
conserved, $N_{1}+N_{2}$, while the ratio $\nu =\left( N_{1}-N_{2}\right)
/\left( N_{1}+N_{2}\right) $, if different from zero, characterizes
spontaneous breaking of symmetry between the components. In fact, $\nu \neq
0 $ is a specific, and most important, manifestation of the immiscibility in
the fermionic system with the linear coupling (in the usual setting, without
the linear interconversion, $N_{1}$ and $N_{2}$ are conserved separately,
which makes the situation altogether different). The VA predicts the onset
of mixing at a critical value of the coupling constant, $\kappa =\kappa _{%
\mathrm{cr}}$, with $\nu \equiv 0$ at $\kappa >\kappa _{\mathrm{cr}}$. The
critical value is predicted as a a function of parameters measuring the
total number of atoms and strength of the nonlinear repulsion between the
species. Comparison with numerical results demonstrates that VA provides a
qualitatively correct description of the transition to miscibility.

Basic numerical results are presented in Sec. III. They demonstrate a
drastic difference of the phase separation in the fermionic mixture from the
immiscible binary BEC: while in the boson gas the two species spontaneously
separate and form a domain wall, with each component occupying the domain on
either side of the wall (both in the ordinary setting, without the linear
coupling \cite{Marek}, and when the coupling is switched on, up to the onset
of the induced mixing \cite{Ilya}). A binary fermionic gas features the
separation in an altogether different form: each component forms a spatially
modulated symmetric (even) density wave, which are anti-locked, with the
phase shift of $\pi $ between the components. A similar picture was
predicted in the fermionic binary gas mixture without the linear coupling
\cite{Sadhan1}. Here, a major novelty is the above-mentioned overall
spontaneous symmetry breaking between the components, manifested by $%
N_{1}/N_{2}\neq 1$ (in the immiscible bosonic condensate, the linear
coupling does not break the equality between $N_{1}$ and $N_{2}$ \cite{Ilya}
). As $\kappa $ approaches $\kappa _{\mathrm{cr}}$ (the mixed state), we
observe $N_{1}/N_{2}\rightarrow 1$, while the modulation depth in the
density waves decreases.

In Sec. IV, we briefly consider additional issues. One of them is the
influence of asymmetry between the species, in the form of a
chemical-potential difference between them, that can be induced by coupling
of the atomic spin to external DC magnetic field. Also considered is a
dynamical relaxation of the binary gas into a mixed state, when $\kappa $ is
suddenly switched from zero to a value exceeding $\kappa _{\mathrm{cr}}$.
The results are summarized in Sec. V.

To conclude the Introduction, it is relevant to mention that the relation
between the imbalance in the spin population, induced by the linear
coupling, and the spatial density-modulation waves developed by the two
interacting species is similar to the known LOFF
(Larkin-Ovchinnikov-Fulde-Ferrell) phase in a mixture of two fermion species
with strongly differing values of the Fermi radius. As shown in original
works \cite{LOFF} (see also review \cite{LOFFreview}), in that situation the
pairing between the fermions belonging to the different species gives rise
to Cooper pairs with a nonzero momentum, which translates into spatial
structures in the form of crystalline patterns of the gap parameter. The
LOFF phase has various implications for quark matter \cite{LOFFreview,quarks}%
, and a possibility of its realization in the mixture of two different spin
states in an ultracold fermion gas, with the interaction between them
controlled via the Feshbach resonance, has been analyzed recently \cite%
{LOFFgas}.

\section{The model and variational approximation}

\subsection{The mean-field hydrodynamic (MFHD)\ equations}

Our starting point is the system of nonlinearly coupled one-dimensional MFHD
equations for the fermion wave functions $\phi $ and $\psi $, as worked out
in Ref. \cite{Sadhan1}, to which we add, following Ref. \cite{Ilya} and
references therein, the linear-coupling terms with strength $\kappa $ and,
in the general case, a chemical-potential difference, $\Delta \mu $ (in the
larger part of the paper, we consider the symmetric system, with $\Delta \mu
=0$). In a scaled form, the equations are:
\begin{eqnarray}
i\phi _{t} &=&-\phi _{xx}+x^{2}\phi +\left( g_{1}|\phi |^{4/3}+g_{2}|\psi
|^{2}\right) \phi -\kappa \psi ,  \label{phi} \\
i\psi _{t} &=&-\psi _{xx}+x^{2}\psi +\left( g_{1}|\psi |^{4/3}+g_{2}|\phi
|^{2}\right) \psi -\kappa \phi   \nonumber \\
&+&\Delta \mu \cdot \psi .  \label{psi}
\end{eqnarray}%
Here, we imply the normalization
\begin{equation}
N=\int_{-\infty }^{+\infty }\left[ \Phi ^{2}(x)+\Psi ^{2}(x)\right] dx\equiv
N_{1}+N_{2}=2.  \label{N}
\end{equation}%
Hence the total number of atoms is hidden in the positive self-repulsion
parameter \cite{Sadhan1}, $g_{1}$, coefficient $g_{2}>0$ accounts for the
nonlinear repulsion between the the species, and the strength of the
trapping potential (the coefficient in front of the parabolic potential, $%
x^{2}$) is scaled to be $1$. The use of the same diagonal nonlinearity
coefficient $g_{1}$ in both equations implies the same number of atoms in
them, and, unless we introduce $\kappa \neq 0$, the number of atoms will
remain equal in both channels. A characteristic feature of the MFHD\
approximation for the fermion gas is the power, $4/3$, of the nonlinear
self-repulsive terms (recall the power is $2$ in the ordinary GPE).

In the particular case of $g_{1}=0$, Eqs. (\ref{phi}) and (\ref{psi}) admit
a different interpretation, being tantamount to a system of GPEs for the
binary BEC \cite{Ilya}, in which the self-interaction coefficient
(intra-species scattering length) is set equal to zero by means of a
Feshbach-resonance adjustment. This particular case establishes a link of
the considered model to those for boson mixtures.

Coefficient $g_{2}$ in Eqs. (\ref{phi}) and (\ref{psi}), which accounts for
the inter-species repulsion, can be controlled by means of the
Feshbach-resonance technique in the fermion binary gas \cite{fsff}. On the
other hand, coefficient $g_{1}$ accounts for the Fermi pressure and cannot
be altered by means of this technique. In the analysis presented below, we
vary both coefficients $g_{1}$ and $g_{2}$, which corresponds to the
normalization adopted in Eqs. (\ref{phi}) and (\ref{psi}) (in particular,
varying $g_{1}$ actually implies the change in the number of atoms in the
gas, as aid above).

We look for stationary solutions, $\left\{ \psi ,\phi \right\} =\left\{ \Psi
(x),\Phi (x)\right\} e^{-i\mu t}$, with chemical potential $\mu $ and real
functions $\Psi $ and $\Phi $ obeying the equations
\begin{eqnarray}
\mu \Phi &=&-\Phi ^{\prime \prime }+x^{2}\Phi +\left( g_{1}\Phi
^{4/3}+g_{2}\Psi ^{2}\right) \Phi -\kappa \Psi ,  \label{Phi} \\
\mu \Psi &=&-\Psi ^{\prime \prime }+x^{2}\Psi +\left( g_{1}\Psi
^{4/3}+g_{2}\Phi ^{2}\right) \Psi -\kappa \Phi  \nonumber \\
&+& \Delta \mu \cdot \Psi ,  \label{Psi}
\end{eqnarray}
(the prime stands for $d/dx$). Together with the normalization condition, $%
\int_{-\infty }^{+\infty }\left[ \Phi ^{2}(x)+\Psi ^{2}(x)\right] dx=2$, as
per Eq. (\ref{N}), Eqs. (\ref{Phi}) and (\ref{Psi}) can be derived as the
variational equations, $\delta L/\delta \Phi =\delta L/\delta \Psi =\partial
L/\partial \mu =0$, from the Lagrangian,
\begin{eqnarray}
L &=&\mu +\frac{1}{2}\int_{-\infty }^{+\infty }\biggr[ -\mu \left( \Phi
^{2}+\Psi ^{2}\right) +\left( \left( \Phi ^{\prime }\right) ^{2}+\left( \Psi
^{\prime }\right) ^{2}\right)  \nonumber \\
& +& x^{2}\left( \Phi ^{2}+\Psi ^{2}\right) +\frac{3}{5}g_{1}\left( \Phi
^{10/3}+\Psi ^{10/3}\right) +g_{2}\Phi ^{2}\Psi ^{2}  \nonumber \\
&-& 2\kappa \Phi \Psi +\frac{1}{2}\Delta \mu \cdot \Psi ^{2}\biggr] dx.
\label{L}
\end{eqnarray}

As said above, spontaneous symmetry breaking between the two species in the
system with the linear coupling, where norms $N_{1}$ and $N_{2}$ are not
conserved separately, can be characterized by the ratio,
\begin{equation}
\nu =\frac{N_{1}-N_{2}}{N_{1}+N_{2}}\equiv \frac{1}{2}\int_{-\infty
}^{+\infty }\left[ \Phi ^{2}(x)-\Psi ^{2}(x)\right] .  \label{nu0}
\end{equation}%
In Eq. (\ref{nu0}), the normalization of $N_{1}+N_{2}$ is taken into account
as per Eq. (\ref{N}). The objective of the analysis is to find $\nu $ as a
function of $\kappa $ (for given $g_{1}$ and $g_{2}$), and thus identify the
above-mentioned critical value, $\kappa _{\mathrm{cr}}$, past which (at $%
\kappa >\kappa _{\mathrm{cr}}$) only the symmetric solution exists, with $%
\nu \equiv 0$ [the symmetric solution exists at $\kappa <\kappa _{\mathrm{cr}%
}$ too, but it is unstable (it does not represent the system's ground state)
in that case, cf. a similar situation in the binary BEC with the linear
coupling \cite{Ilya}].

\subsection{Variational approximation}

The only tractable \textit{ansatz} with which the VA can be applied to
Lagrangian (\ref{L}) is based on simple Gaussians,
\begin{equation}
\Phi =ae^{-x^{2}/2},~\Psi =be^{-x^{2}/2},  \label{primitive}
\end{equation}
with constants $a$ and $b$. Normalization condition (\ref{N}) for this
ansatz takes the form of
\begin{equation}
\sqrt{\pi }\left( a^{2}+b^{2}\right) =2,  \label{Nprimitive}
\end{equation}
and the symmetry-breaking parameter, defined by Eq. (\ref{nu0}), is
\begin{equation}
\nu =\left( \sqrt{\pi }/2\right) \left( a^{2}-b^{2}\right) .  \label{nu}
\end{equation}

Following the general method of the VA \cite{VA}, we substitute ansatz (\ref%
{primitive}) in Lagrangian (\ref{L}) and perform the integration, which
yields
\begin{eqnarray}
\frac{L}{\sqrt{\pi }} &=&\frac{\mu }{\sqrt{\pi }}+\frac{1-\mu }{2}\left(
a^{2}+b^{2}\right) +\frac{3\sqrt{15}g_{1}}{50}\left( a^{10/3}+b^{10/3}\right)
\nonumber \\
&+&\frac{g_{2}}{2\sqrt{2}}\left( ab\right) ^{2}~-\kappa ab.
\label{improvedL}
\end{eqnarray}%
(normalization condition (\ref{Nprimitive}) can also be derived from this
Lagrangian, as $\partial L/\partial \mu =0$). In the framework of the VA,
constants $a$ and $b$ are determined by equations $\partial L/\partial
a=\partial L/\partial b=0$, i.e.,
\begin{eqnarray*}
\sqrt{\frac{3}{5}}g_{1}a^{7/3}\allowbreak +\frac{1}{\sqrt{2}}%
g_{2}ab^{2}-\kappa b &=&\left( \mu -1\right) a, \\
\sqrt{\frac{3}{5}}g_{1}b^{7/3}\allowbreak +\frac{1}{\sqrt{2}}%
g_{2}ba^{2}-\kappa a &=&\left( \mu -1\right) b.
\end{eqnarray*}%
Eliminating $\mu $\ from these equations leads to a relation in which $(a-b)$
factorizes out, which yields an obvious symmetric solution, $a=b$. The
remaining equation which determines the asymmetric solutions is
\begin{eqnarray}
\sqrt{\frac{3}{5}} &g_{1}&ab\left( a^{2/3}+b^{2/3}\right) -\left( \frac{g_{2}%
}{\sqrt{2}}ab-\kappa \right)  \nonumber \\
&\times &\left[ a^{4/3}+b^{4/3}+\left( ab\right) ^{2/3}\right] =0.
\label{improved}
\end{eqnarray}

Equations (\ref{Nprimitive}) and (\ref{improved}) constitute a system from
which $a$ and $b$ should be found, for given $g_{1}$ and $g_{2}$. Although
this system cannot be solved in an analytical form, the critical value, $%
\kappa _{\mathrm{cr}}$, can be easily found. Indeed, since only the
symmetric solution exists at $\kappa \geq $ $\kappa _{\mathrm{cr}}$, the
respective \textit{bifurcation}, at which the asymmetric solution merges
into the symmetric one and disappears, occurs when Eqs. (\ref{Nprimitive})
and (\ref{improved}) themselves admit the symmetric solution, $a=b=\pi
^{-1/4}$. The substitution of this in Eq. (\ref{improved}) yields a simple
result,
\begin{equation}
\kappa _{\mathrm{cr}}\equiv \frac{g_{2}}{\sqrt{2\pi }}-\frac{2g_{1}}{\sqrt{%
15 }\pi ^{1/3}}~.  \label{new-cr}
\end{equation}
Note that Eq. (\ref{new-cr}) shows that $\kappa _{\mathrm{cr}}$ vanishes at
the point
\begin{equation}
g_{2}=g_{2}^{\mathrm{cr}}\equiv 2\sqrt{\frac{2}{15}}\pi ^{1/6}g_{1}\approx
0.884g_{1}.  \label{new-gcr}
\end{equation}
The meaning of the latter result is that, for $g_{2}<g_{2}^{\mathrm{cr}}$,
the VA predicts the binary fermionic gas to be in the mixed state even
without the linear-coupling terms.

\begin{figure}[tbp]
\par
\includegraphics[width=\linewidth]{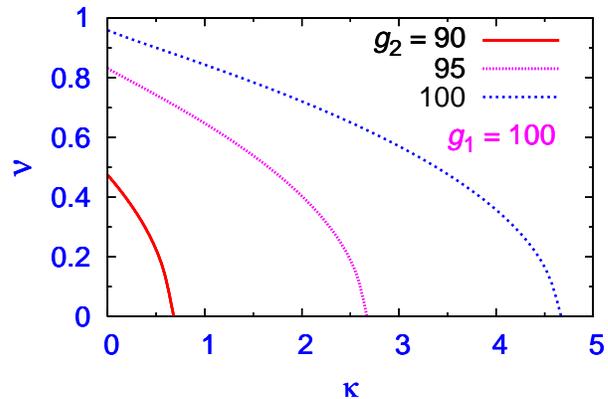}
\caption{(Color online) A set of bifurcation diagrams $-$ $\protect\nu $ vs.
$\protect\kappa $ plots $-$ which show [in terms of the number-of-atoms
symmetry-breaking parameter, $\protect\nu $, see Eq. (\protect\ref{nu})] the
asymmetric solution, as found from the variational equations, (\protect\ref%
{Nprimitive}) and (\protect\ref{improved}), for a fixed $g_{1}$ and varying
values of $g_{2}$. The solution with $\protect\nu =0$ is an obvious
symmetric one, $a=b=\protect\pi ^{-1/4}$, see text.}
\label{fig:bif}
\end{figure}

Close to the bifurcation which takes place at $\kappa =\kappa _{\mathrm{cr}}$
, i.e., for $0<\Delta \kappa \equiv \kappa _{\mathrm{cr}}-\kappa \ll \kappa
_{\mathrm{cr}}$, the system of equations (\ref{Nprimitive}) and (\ref%
{improved}) can be expanded, which yields the corresponding slightly
asymmetric solution,
\begin{equation}
a=\pi ^{-1/4}\pm \Delta a,~b=\pi ^{-1/4}\mp \Delta a,  \label{+-}
\end{equation}%
where, in the first approximation,
\begin{equation}
\left( \Delta a\right) ^{2}=\Delta \kappa \cdot \frac{27\sqrt{30}}{31\sqrt{15%
}g_{2}-56\sqrt{2}\pi ^{1/6}g_{1}}.  \label{Delta}
\end{equation}%
It is relevant to note that the condition necessary for the symmetry
breaking (within the framework of the VA), $g_{2}>g_{2}^{\mathrm{cr}}$, see
Eq. (\ref{new-gcr}), guarantees that Eq. (\ref{Delta}) yields $\left( \Delta
a\right) ^{2}>0$ for $\Delta \kappa >0$, as it must be. Farther from the
bifurcation point, Eqs. (\ref{Nprimitive}) and (\ref{improved}) can be
solved numerically, which gives rise to a typical bifurcation picture
displayed in Fig. \ref{fig:bif}, where we plot $\nu $ vs. $\kappa $ for $%
g_{1}=100$ and $g_{2}=90$, $95$, and $100$. In this case, the fermionic gas
remains in the mixed state for $g_{2}<g_{2}^{\mathrm{cr}}\approx 88.4$ even
for $\kappa =0$. For $g_{2}>g_{2}^{\mathrm{cr}}$, the two species separate,
and the separation increases with growing $g_{2}$. Nevertheless, from Fig. %
\ref{fig:bif} we find that, for $g_{2}=90$, $95$, and $100$, parameter $\nu $
vanishes for $\kappa >0.68$ ,$2.67$, and $4.67$, respectively, and the
system becomes mixed again. Comparison of further analytical predictions
with numerical results are presented in detail in the next section.

\section{Numerical results}

We solved coupled MFHD equations (\ref{phi}) and (\ref{psi}) numerically,
using a real-time integration method based on the Crank-Nicholson
discretization scheme, as elaborated in Ref. \cite{sk1}. We discretize the
mean-field equations, using time step $0.001$ and space step $0.05$, in
domain $-25<x<25$. We start with the Gaussian ground-state wave function of
the linear harmonic oscillator as the input at $t=0$, $\Phi (x,t=0)=\Psi
(x,t=0)=\pi ^{-1/4}\exp (-x^{2}/2)$, while setting $g_{1}=g_{2}=0$. With
these initial states, we have simulated the time evolution of Eqs. (\ref{phi}%
) and (\ref{psi}). In the course of the evolution, nonlinearity coefficients
$g_{1}$ and $g_{2}$ are gradually switched on, at the rate of 0.0001 per
time step of 0.001. Solutions attain the final stationary configuration
after the full nonlinearities and linear coupling $\kappa $ have been
restored. All the wave functions displayed below are definitely stable
against small perturbations, as stationary solutions to Eqs. (\ref{phi}) and
(\ref{psi}).

With $\kappa=0$, the norms of the two components are conserved separately.
In that case, we also used an imaginary-time-integration method based on the
Crank-Nicholson discretization scheme, and the results obtained for Eqs. (%
\ref{phi}) and (\ref{psi}) are in agreement with the integration in real
time, which verifies the correctness of the solutions. However, with $%
\kappa\ne 0$, the normalizations of the two components are not separately
conserved, and the imaginary-time-iteration method is not applicable.

\begin{figure}[tbp]
\begin{center}
\includegraphics[width=\linewidth]{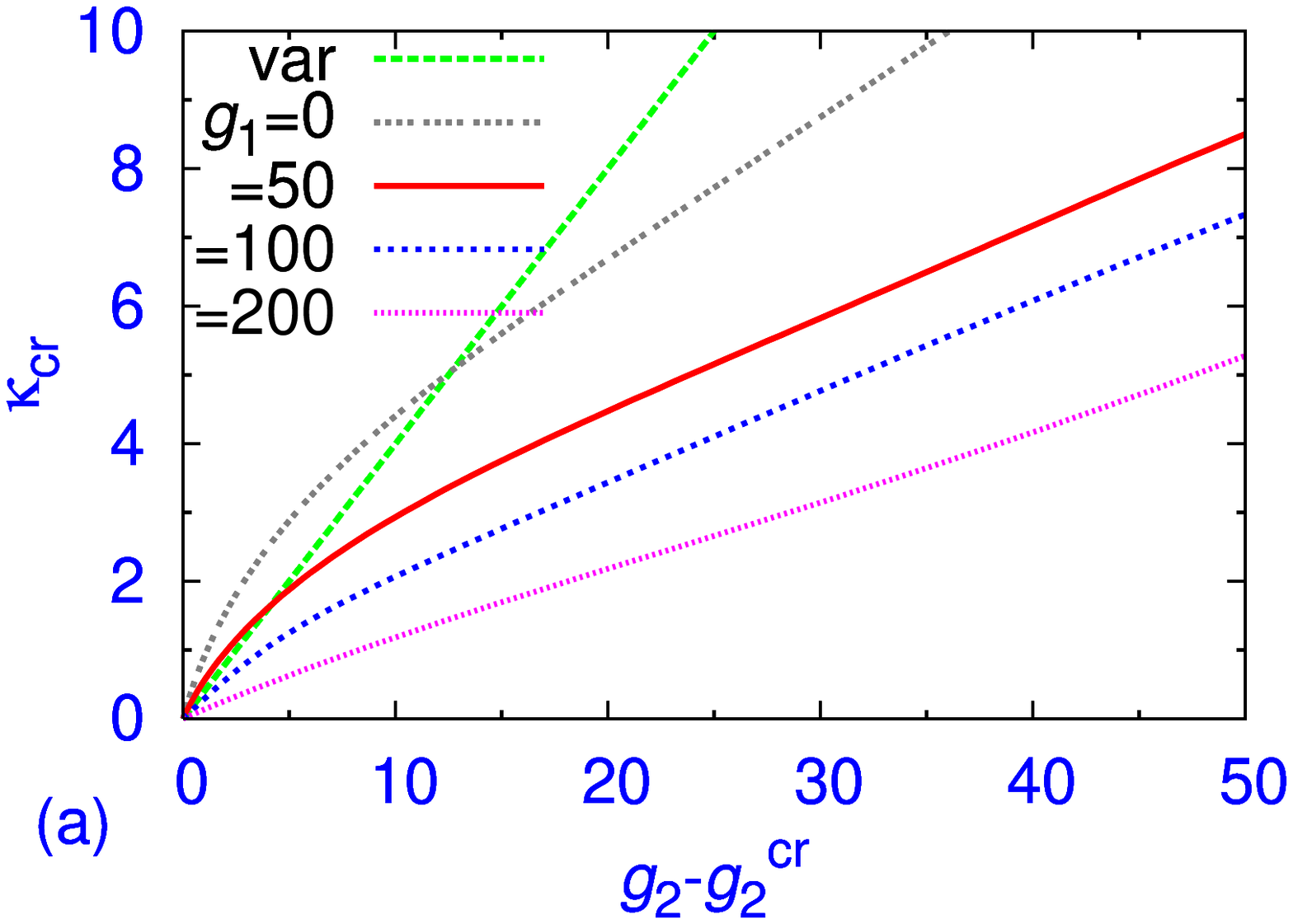} 
\includegraphics[width=\linewidth]{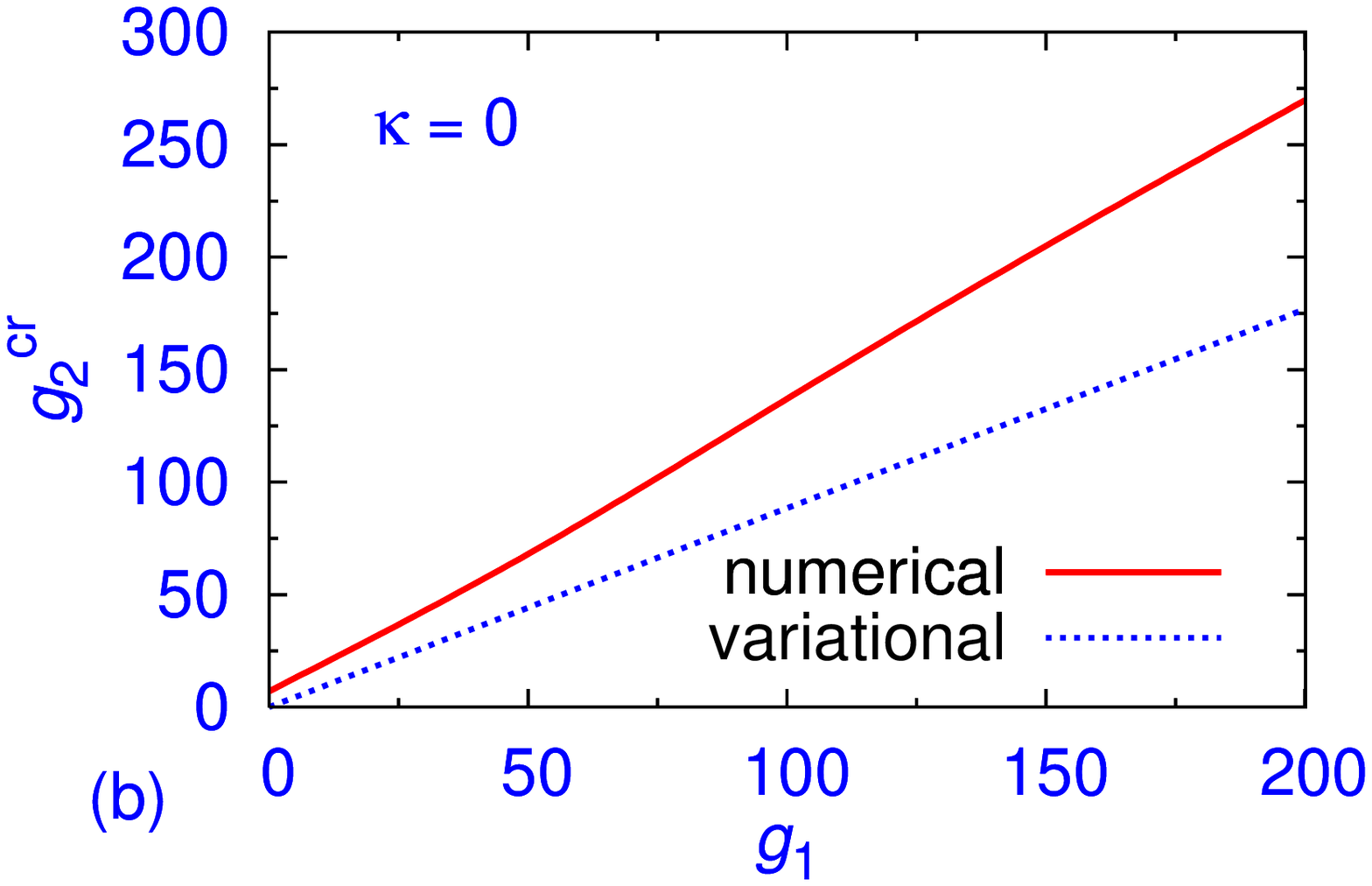} 
\end{center}
\caption{(Color online) (a) The dependence of the critical value of the
linear-coupling strength $\protect\kappa _{\mathrm{cr}}$ on the
nonlinear-repulsion constant $g_{2}$, plotted as $\protect\kappa _{\mathrm{cr%
}}$ vs. ($g_{2}-g_{2}^{\mathrm{cr}}$) for $g_{1}=0,50,100,$ and 200.
Numerically obtained curves are labeled by respective values of $g_{1}$,
while their counterpart produced by the variational approximation
variational is labeled ``var". Recall $g_{2}^{\mathrm{cr}}$ is the value of $%
g_{2}$ below which the binary gas remains completely mixed in the absence of
the linear coupling. (b) The numerically found dependence of $g_{2}^{\mathrm{%
cr}}(g_{1})$, along with the respective variational prediction given by Eq. (%
\protect\ref{new-gcr}). The numerical dependence is (almost exactly) linear,
although the slope, $1.3$, is somewhat different from its variational
counterpart, $0.884$.}
\label{fig:compare}
\end{figure}

A comparison of the analytical (variational) predictions from the previous
section with numerical results can be made in terms of $\kappa _{\mathrm{cr}%
} $ as a function of $g_{2}-g_{2}^{\mathrm{cr}}$, for fixed values of $g_{1}$%
. This comparison is displayed in Fig. \ref{fig:compare}(a), where we plot
variational and numerical results for $g_{1}=0$, $50$, $100$, and $200$.
This plot demonstrates that the simple VA provides for a qualitatively
reasonable prediction, provided that $g_{2}-g_{2}^{\mathrm{cr}}$ and $g_{1}$
are not too large.

A simple but noteworthy feature of the VA is the linear dependence of $%
g_{2}^{\mathrm{cr}}$ on $g_{1}$, see Eq. (\ref{new-gcr}). It can be compared
to its numerically found counterpart, as shown in Fig. \ref{fig:compare}(b).
The numerical dependence is nearly linear too, with the slope $dg_{2}^{%
\mathrm{cr}}/dg_{1}\approx 1.3$. The comparison with the analytical
prediction for the slope, which is $0.884$ as per Eq. (\ref{new-gcr}), shows
that the VA yields qualitatively correct, but not quite accurate, results.
For $g_{1}=0$, the VA yields $g_{2}^{\mathrm{cr}}=0$, whereas the
corresponding numerical result is $g_{2}^{\mathrm{cr}}\approx 7$. This means
that, even for binary atomic systems with no intraspecies interaction ($%
g_{1}=0$), inter-species interaction $g_{2}$ has to be above the critical
value $g_{2}^{\mathrm{cr}}\approx 7$ for the demixing to take place (which
is a consequence of the fact that the two species can be mixed due to the
pressure from the trapping potential). The simple VA fails to predict this
fact, allowing for the demixing at any $g_{2}^{\mathrm{cr}}$.

Besides the conspicuous error in predicting the slope of the $g_{2}^{\mathrm{%
cr}}(g_{1})$ dependence, another shortcoming of the VA is seen in the fact
that the analytically predicted slope of the dependence of $\kappa _{\mathrm{%
cr}}$ vs. $g_{2}-g_{2}^{\mathrm{cr}}$ does not depend on $g_{1}$, while the
numerically found slope changes with $g_{1}$ (in this connection, it is
relevant to recall that the limit case of $g_{1}=0$, which is included in
Fig. \ref{fig:compare} (a), corresponds, as explained above, not to the MFHD
approximation for the fermion mixture, but rather to the system of coupled
GPEs for the binary boson condensate with zero intra-species interaction).

The VA could, in principle, be improved. An essential inaccuracy of simple
ansatz (\ref{primitive}) is the assumption that the wave functions
monotonically decrease with $|x|$; in reality, numerically found wave
functions of the ground state have several local minima and maxima, see Fig. %
\ref{fig:examples} below. Even in the fully mixed state, the profile may be
markedly different from the Gaussian, as suggested by the curves pertaining
to $\kappa =1.5$ in that figure. The simplest possibility to improve ansatz (%
\ref{primitive}) in this respect would be to replace it by a generalized
one, $\Phi =\left( a_{0}+a_{2}x^{2}\right) e^{-x^{2}/2},~\Psi =\left(
b_{0}+b_{2}x^{2}\right) e^{-x^{2}/2}$, with extra constants $a_{2},b_{2}$.
However, although the Lagrangian corresponding to the extended ansatz can be
calculated in an analytical form, the variational equations following from
it are so cumbersome that direct numerical solution of exact equations (\ref%
{phi}) and (\ref{psi}), or equivalently, (\ref{Phi}) and (\ref{Psi}), is
actually simpler.


Now we consider the results for the symmetric system $\Delta \mu =0$.
Collecting the results for many values of $g_{1}$ and $g_{2}$, we have
concluded that details of the solutions may differ considerably, but their
qualitative properties are universal. An appropriate case to display generic
results is one with $g_{1}=142$, $g_{2}=100$ (in this case, $g_{2}^{\mathrm{%
cr}}=137$). In Fig. \ref{fig:examples}, we display a set of ground-state
wave functions found as $\kappa $ increases towards the transition to the
full mixing at $\kappa =\kappa _{\mathrm{cr}}$ (in this case, $\kappa _{%
\mathrm{cr}}\approx 1.50$). A salient (and truly generic, as shown by
comparison with a large pool of numerical data obtained at other values of $%
g_{1}$ and $g_{2}$) feature of the ground state in the immiscible phase is
that the two species do not collect themselves in different spatial domains,
separated by a spontaneously formed wall (which is the case in the boson
mixtures \cite{Marek,Ilya}), but, instead, they form two spatially modulated
\textit{anti-locked} density waves (i.e., ones with a phase shift of $\pi $%
), while each wave remains an even function of $x$. In the absence of the
linear coupling, a qualitatively similar structure of the ground state of
the binary fermionic gas, with equal numbers of atoms in the two components,
was reported in Ref. \cite{Sadhan1}. As $\kappa $ approaches the miscibility
point, $\kappa _{\mathrm{cr}}$, the modulation in both components becomes
more shallow, and finally completely disappears for $\kappa \geq \kappa _{%
\mathrm{cr}}$, where a full mixing between the components takes place.

\begin{figure}[tbp]
\par
\begin{center}
\includegraphics[width=\linewidth]{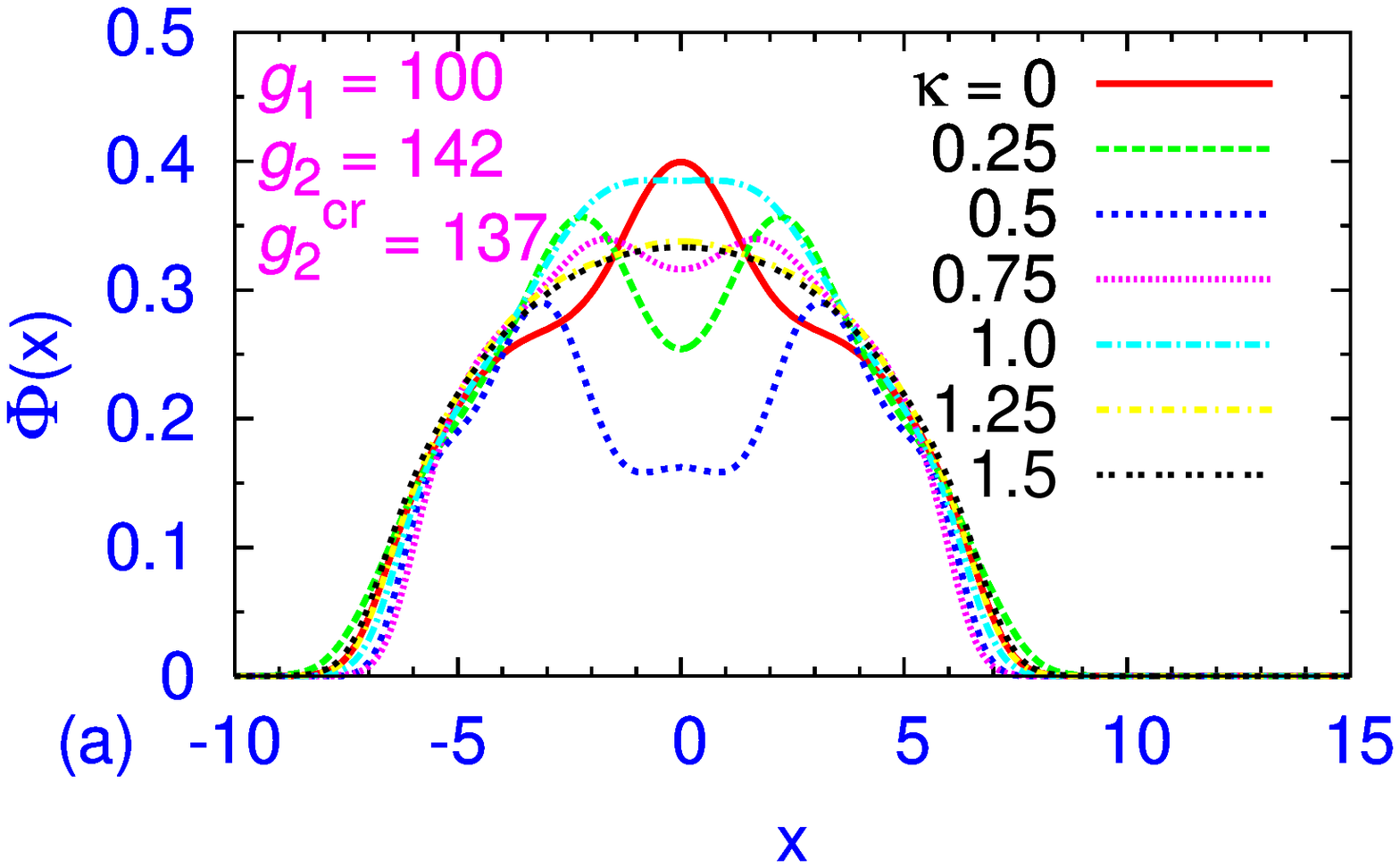} 
\includegraphics[width=\linewidth]{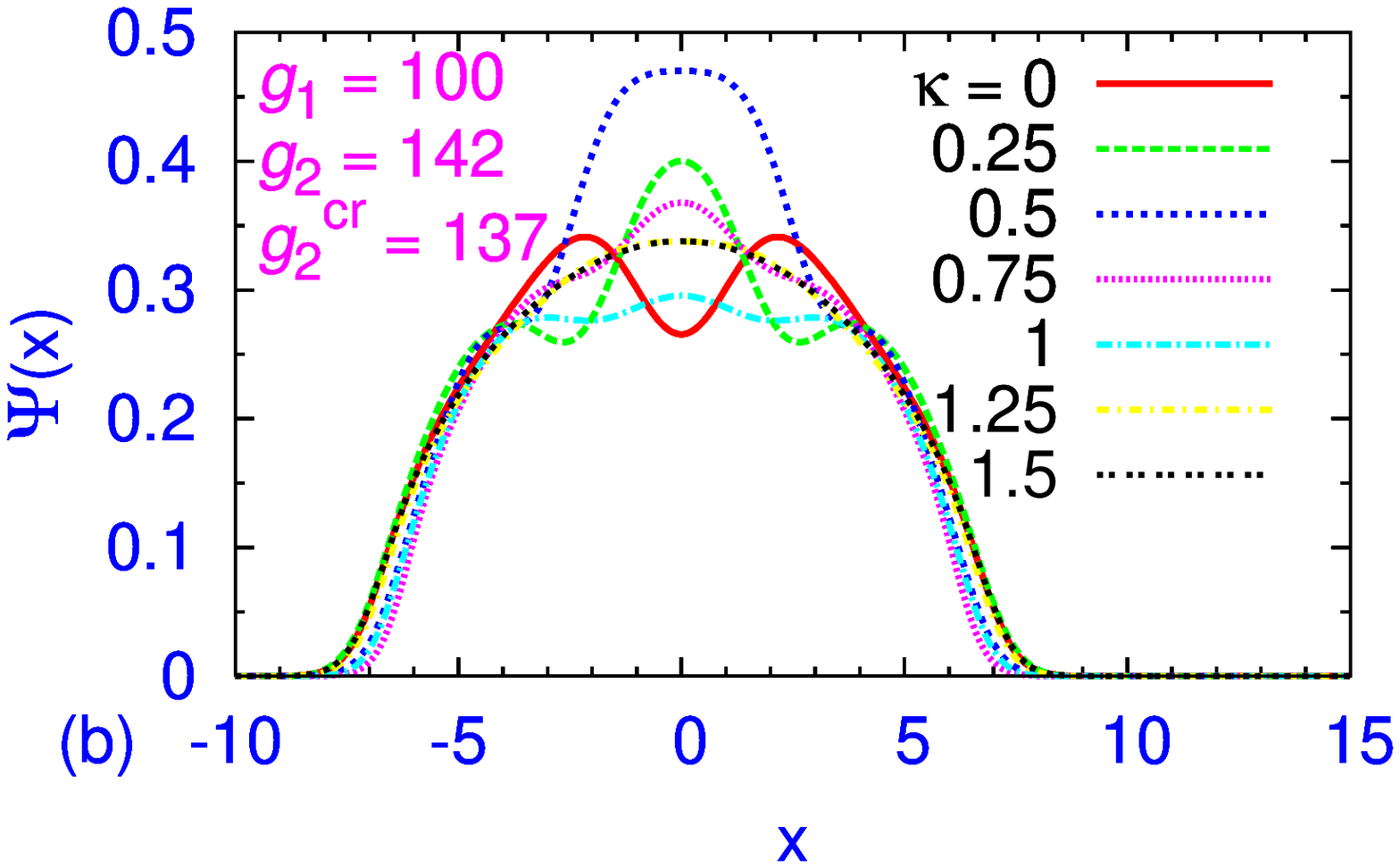} 
\end{center}
\caption{(Color online). A generic example of a set of ground-state wave
functions of the binary mixture, $\Phi (x)$ and $\Psi (x)$, as obtained from
the numerical solution of Eqs. (\protect\ref{phi}) and (\protect\ref{psi})
in the symmetric case ($\Delta \protect\mu =0$) for $g_{1}=100$ and $%
g_{2}=142$ and varying $\protect\kappa $. }
\label{fig:examples}
\end{figure}

As overall characteristics of the immiscible state, we use the
number-of-atoms symmetry-breaking parameter, defined by Eq. (\ref{nu0}),
and, in addition,
\begin{equation}
\Xi \equiv \frac{\int_{-\infty }^{+\infty }x^{2}\left[ \Phi ^{2}(x)-\Psi
^{2}(x)\right] dx}{\int_{-\infty }^{+\infty }x^{2}\left[ \Phi ^{2}(x)+\Psi
^{2}(x)\right] dx},  \label{Xi}
\end{equation}%
which quantifies the anti-locking between the modulated density waves in the
two components. For the same typical case which is illustrated by the set of
ground-state profiles in Fig. \ref{fig:examples}, $\nu $ and $\Xi $ are
displayed, as functions of $\kappa $, in Fig. 4. A salient feature of the
dependences (which is another truly generic feature, observed in all other
cases for which numerical solutions were obtained) is the oscillatory
character of the decay of both $\nu $ and $\Xi $, observed as $\kappa $
approaches $\kappa _{\mathrm{cr}}$. Naturally, we have $\nu =0$ at $\kappa =0
$ (we always started the computation from a symmetric state, with $%
N_{1}=N_{2}$, at $\kappa =0$, when the two numbers of atoms are conserved
separately), while $\Xi (\kappa =0)$ is different from zero, because, as
mentioned above, the anti-locking between the modulated density waves in the
two components of the ground state (i.e., \textit{demixing} in this state)
takes place also in the system without the linear interconversion \cite%
{Sadhan1}. Because the symmetry breaking between the two components is
spontaneous, the overall signs of $\nu $ and $\Xi $ are, of course, selected
arbitrarily. The plots in Fig. 4 are generated by choosing $\Xi (\kappa =0)>0
$, and then drawing the curves by continuity.

\begin{figure}[tbp]
\begin{center}
\includegraphics[width=\linewidth]{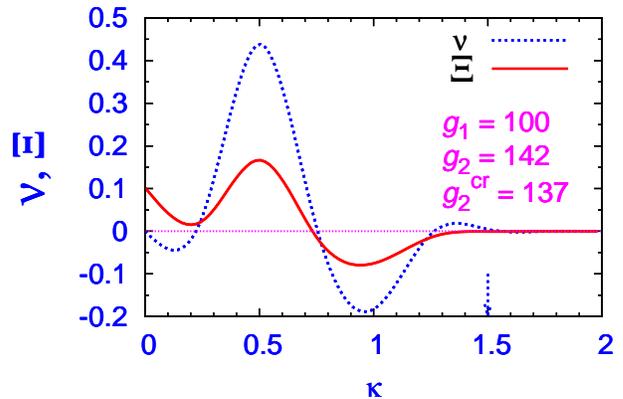}
\end{center}
\par
\label{fig:nu&Xi}
\caption{(Color online) Parameters which characterize the spontaneously
established inequality in numbers of atoms between the linearly coupled
immiscible species, $\protect\nu $ [see Eq. (\protect\ref{nu0})], and
anti-locking in the spatial density modulation of the species, $\Xi $ [see
Eq. (\protect\ref{Xi})], are shown as functions of $\protect\kappa $, found
from the numerical solution of Eqs. (\protect\ref{phi}) and (\protect\ref%
{psi}) in the symmetric case ($\Delta \protect\mu =0$). The critical value $%
\protect\kappa _{\mathrm{cr}}\approx 1.50$ of $\protect\kappa $ above which
complete mixing takes place in this case is shown by an arrow.}
\end{figure}

\begin{figure}[tbp]
\begin{center}
\includegraphics[width=\linewidth]{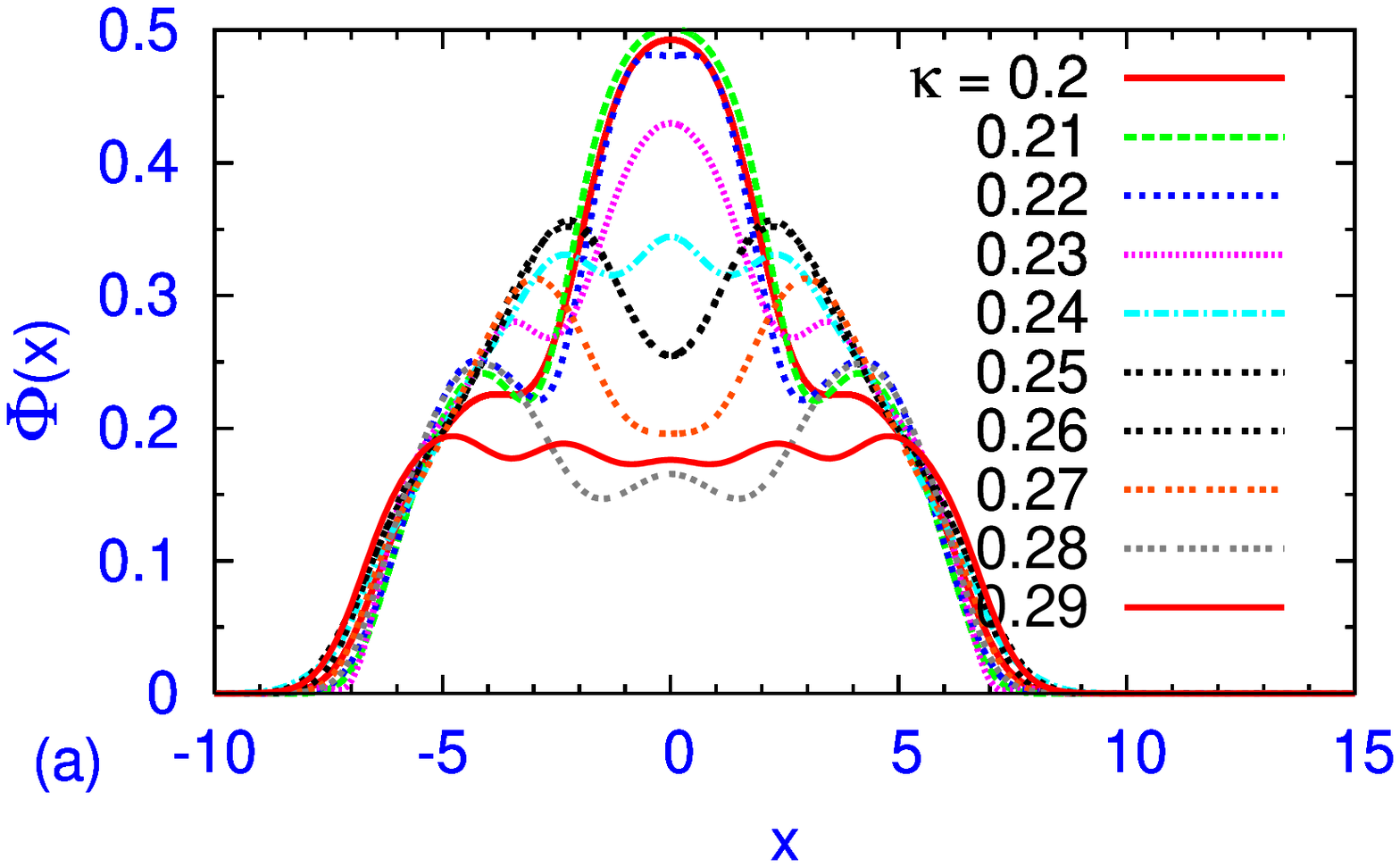} 
\includegraphics[width=\linewidth]{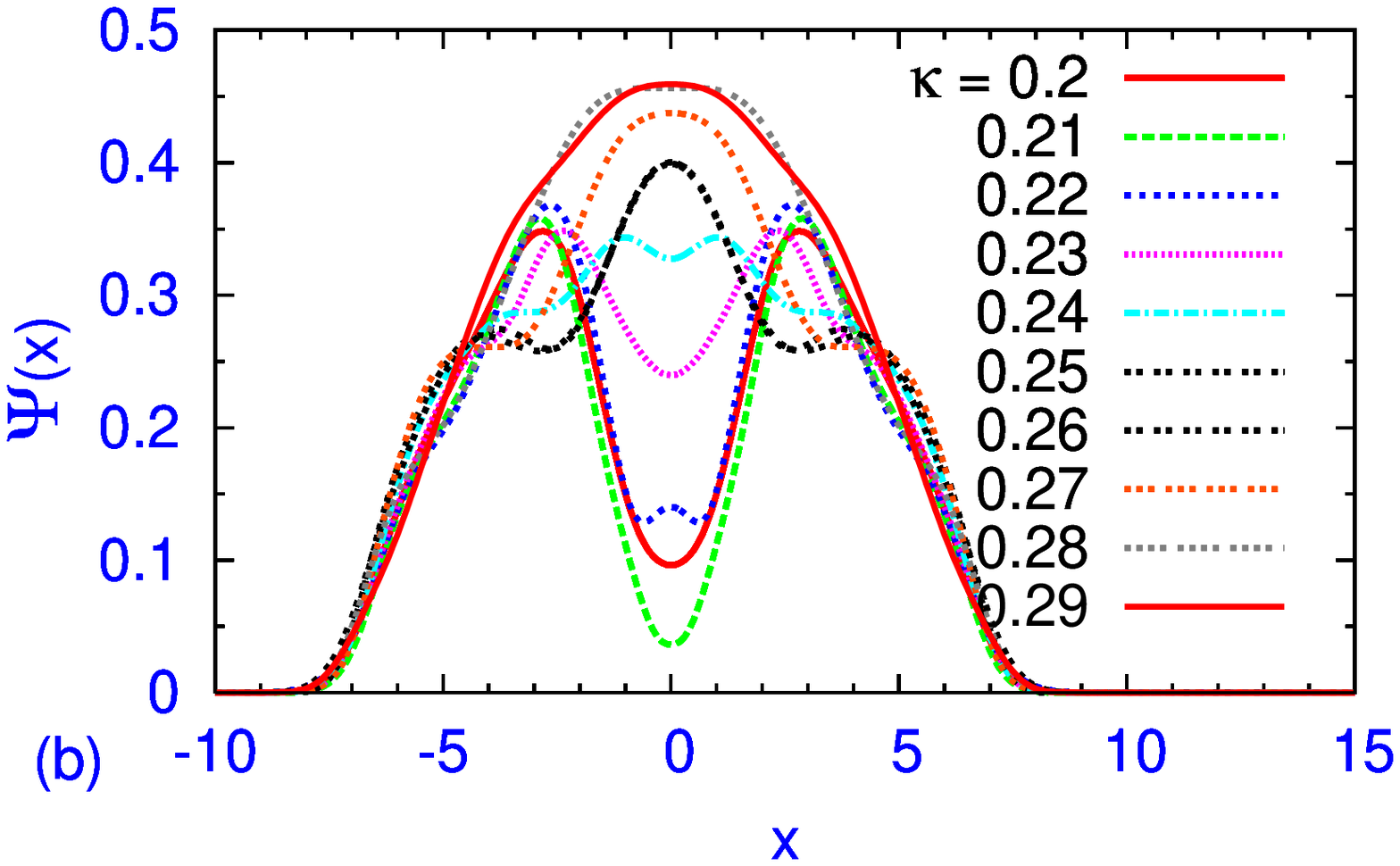} 
\end{center}
\caption{(Color online). In addition to Fig. \protect\ref{fig:examples}, an
extra set of profiles of the wave functions in the symmetric case ($\Delta
\protect\mu =0$), for $g_{1}=100$ and $g_{2}=142$, in the ground state
around $\protect\kappa =0.25$, where nearly complete mixing occurs
accidentally. }
\label{fig:extra-profiles}
\end{figure}

Another feature that deserves special consideration is that, at some values
of $\kappa $ (in Fig. 4, these are $\kappa \approx 0.25$ and $\kappa \approx
0.75$), both $\nu $ and $\Xi $ take nearly zero values almost
simultaneously, suggesting that the system falls into a nearly mixed state
at these points. Indeed, the consideration of the profiles in a vicinity of
one of these points, see Fig. \ref{fig:extra-profiles}, shows that the two
profiles are almost identical at $\kappa $ close to $0.24$; nevertheless,
the modulation does not completely disappear in this case, and the situation
seems as an accidental mixing. As $\kappa $ moves away from this point, both
$\nu $ and $\Xi $ become different from zero, which implies restoration of
the demixing and difference in numbers of atoms in the two components, as
seen in Fig. \ref{fig:extra-profiles}.

It is noteworthy that the deepest anti-locked modulation in the two
components (in other words, a maximum of $|\Xi (\kappa )|$) is attained not
at $\kappa =0$, but at a finite value of the coupling constant. As seen from
Figs. \ref{fig:examples} and 4, in the present example this value is $\kappa
\approx 0.5$.

\section{Further results}

\subsection{The asymmetric system ($\Delta \protect\mu \neq 0$)}

We now proceed to the role of the chemical-potential asymmetry in Eqs. (\ref%
{phi}) and (\ref{psi}). If the asymmetry is present in Eq. (\ref{psi}), $%
\Delta \mu \neq 0$, the inequality of the numbers of atoms in the asymmetric
components is not spontaneous (i.e., the sign of $\nu $ is not selected
arbitrarily), unlike the symmetric system: for $\Delta \mu >0$, $N_{1}$ is
larger than $N_{2}$ (similar to what was found in the model of the
asymmetric boson mixture \cite{Ilya}). Accordingly, $\Delta \mu $ tends to
suppress the oscillations in dependences $\nu (\kappa )$ and $\Xi (\kappa )$%
, which were a salient feature of the symmetric system, see Fig. 4. Our
results clearly show (Fig. 7) that, in the model with a nonzero difference
in the chemical potentials, $N_1$ and $N_2$ are never exactly equal.

\begin{figure}[tbp]
\begin{center}
\includegraphics[width=\linewidth]{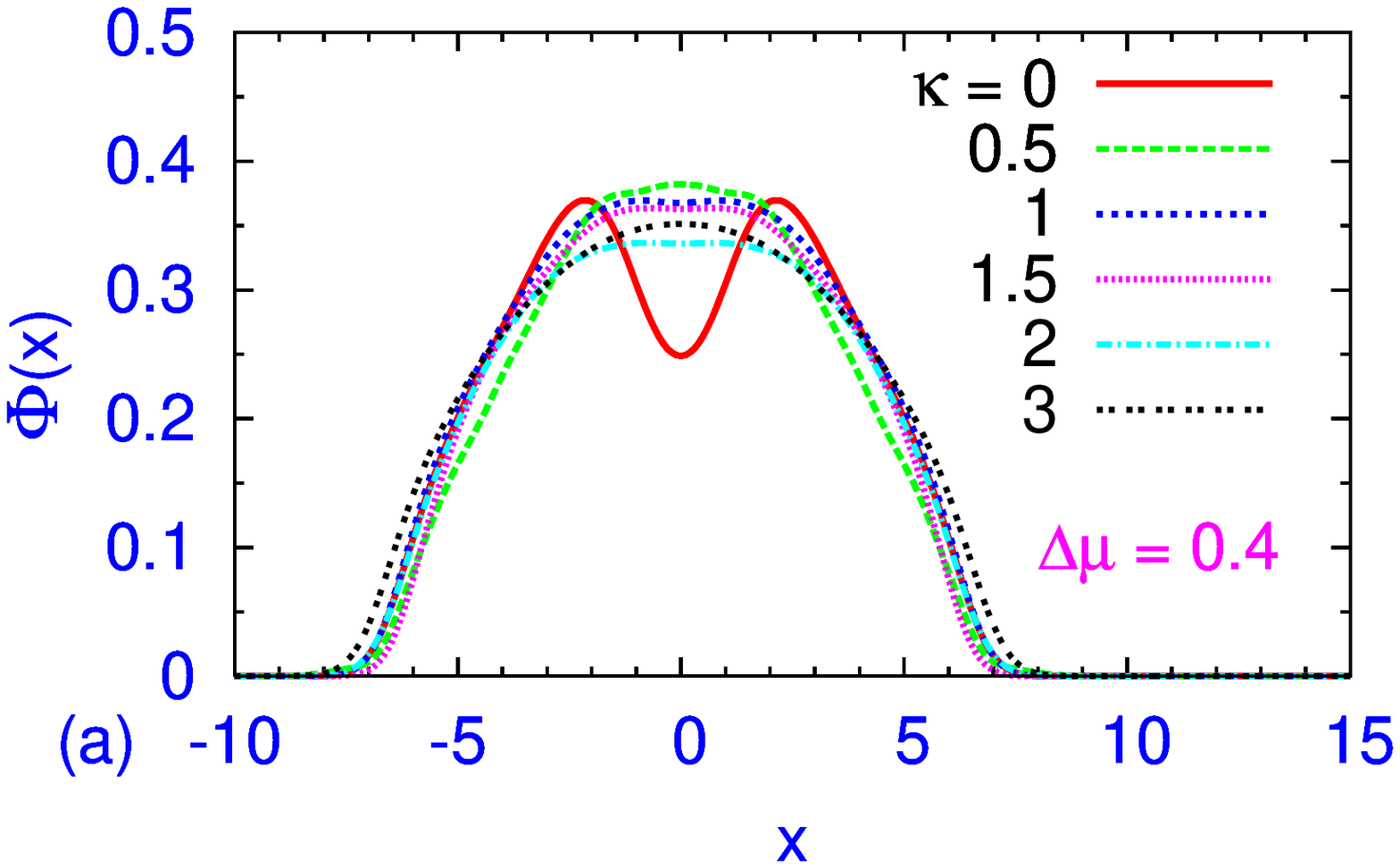} 
\includegraphics[width=\linewidth]{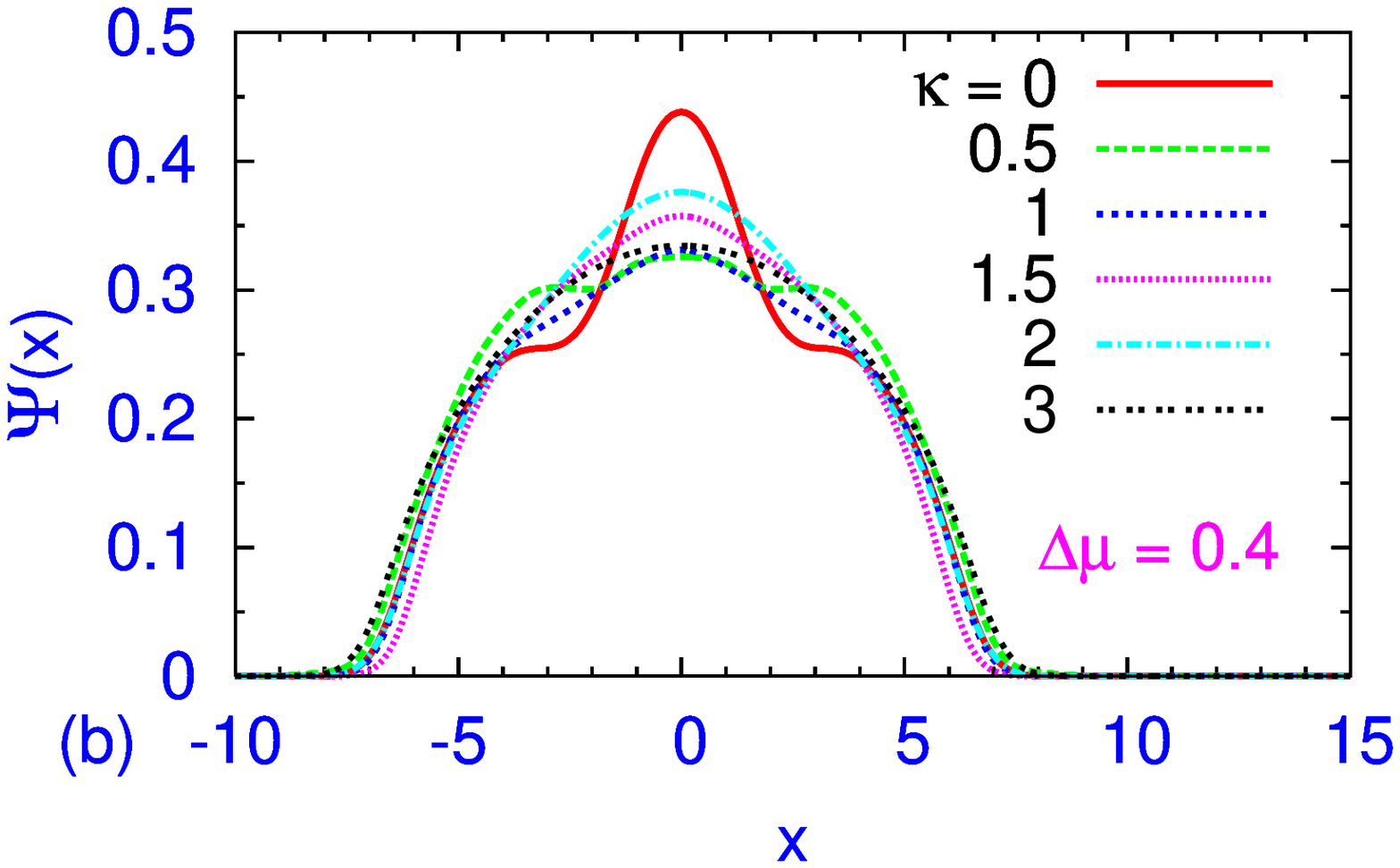} 
\end{center}
\par
\caption{(Color online). A generic example of a set of ground-state wave
functions of the binary mixture, $\Phi (x)$ and $\Psi (x)$, for different $%
\protect\kappa $, as obtained from numerical solution of Eqs. (\protect\ref%
{phi}) and (\protect\ref{psi}) in the asymmetric model, with $g_{1}=100$, $%
g_{2}=142$ and $\Delta \protect\mu =0.4.$ }
\label{fig:examples-asymm}
\end{figure}

In Fig. \ref{fig:examples-asymm}, we display a counterpart of Figs. \ref%
{fig:examples} and \ref{fig:extra-profiles}, i.e., a set of profiles of the
wave functions in the ground state for $g_{1}=100$ and $g_{2}=142$, in the
presence of relatively strong asymmetry, with $\Delta \mu =0.4$. Some
residual oscillations in the shape of the wave functions are still seen, but
they are much less pronounced than in Figs. \ref{fig:examples} and \ref%
{fig:extra-profiles}.

\begin{figure}[tbp]
\includegraphics[width=\linewidth]{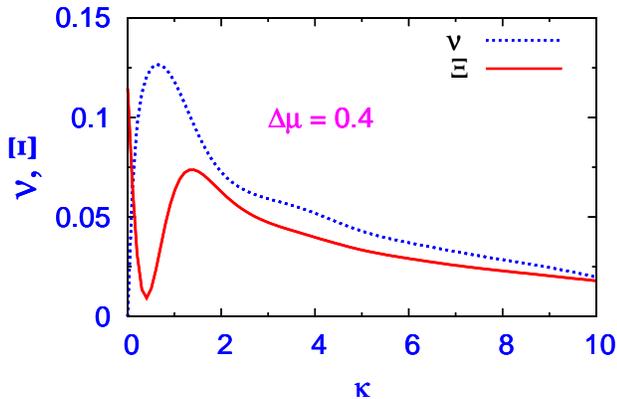}
\caption{(Color online) The same as in Fig. 4, but for the asymmetric system
corresponding to Fig. \protect\ref{fig:examples-asymm}, i.e., with $\Delta
\protect\mu =0.4$ in Eq. (\protect\ref{psi}).}
\label{fig:nu&Xi-asymm}
\end{figure}

A counterpart of Fig. 4 for the same example of the asymmetric system is
displayed in Fig. \ref{fig:nu&Xi-asymm}, where we plot $\nu $ and $\Xi $ vs.
$\kappa $ for the asymmetric fermion mixture studied in Fig. \ref%
{fig:examples-asymm}. As expected, the figure shows that the oscillations in
the plots of $\nu (\kappa )$ and $\Xi (\kappa )$ are greatly reduced, and,
on the contrary to the symmetric model (see Fig. 4), there is no point at
which these functions, $\nu (k)$ and $\Xi (k)$, would vanish; instead, they
decay asymptotically at large values of $\kappa $. The nonvanishing of $\nu
(k)$ implies that the numbers of atoms in the two components never become
exactly equal at $\Delta \mu \neq 0$.

\subsection{Dynamical relaxation}

\begin{figure}[tbp]
\includegraphics[width=\linewidth]{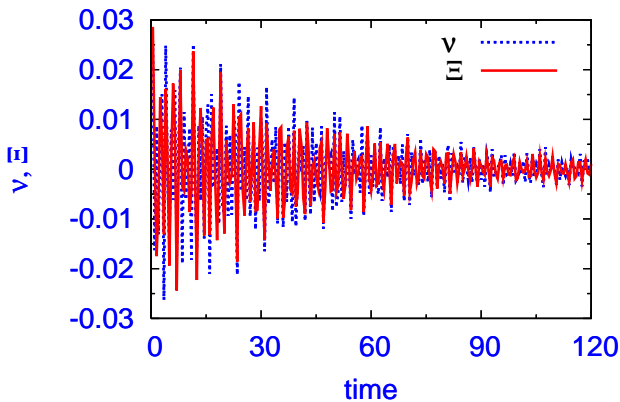}
\caption{(Color online) Relaxation of the initial immiscible state in the
system with $g_{1}=100,g_{2}=140$ and $\protect\kappa =0$ into a mixed
state, after $\protect\kappa $ was suddenly switched on from $0$ to $4$
(recall that $\protect\kappa _{\mathrm{cr}}\approx 1.5$ for these values of $%
g_{1}$ and $g_{2}$).}
\label{fig:dynamical}
\end{figure}

The above analysis was based on stationary solutions, found from Eqs. (\ref%
{Phi}) and (\ref{Psi}) [the stability of the solutions was verified within
the framework of the full time-dependent MFHD equations, (\ref{phi}) and (%
\ref{psi})]. Another aspect of the immiscibility/miscibility transition is
relaxation of an initial immiscible state after the coupling terms were
suddenly switched on. In the context of the boson mixture, a similar problem
was considered in Ref. \cite{Ilya}. In that work, it was found that the
sudden application of the strong linear coupling did not lead to relaxation
of the system into a mixed state, but rather to persistent nonequilibrium
oscillations. Here, we performed numerical experiments, starting from the
phase-separated state at $\kappa =0$ and suddenly adding the linear coupling
between the components, with $\kappa >\kappa _{\mathrm{cr}}$ (in the
symmetric system, with $\Delta \mu =0$). The result can be adequately
presented in the form of the time dependence of the immiscibility
parameters, $\nu $ and $\Xi $, that were defined above [see Eqs. (\ref{nu0})
and (\ref{Xi})]. A typical example of this dependence is displayed in Fig. %
\ref{fig:dynamical}, which shows oscillatory relaxation to the mixed state
characterized by $\nu =\Xi =0$.

\section{Conclusion}

We have demonstrated that the linear coupling between two different spin
states (induced by a resonant electromagnetic field), which form an
immiscible binary fermionic gas, may enforce a transition to miscibility.
The description of the degenerate gas is based on a system of equations for
the wave functions of the two components, derived in the MFHD\
(mean-field-hydrodynamic) approximation. The transition to the induced
miscibility was predicted by means of the variational approximation, and
then verified in numerical computations. In drastic contrast with the
induced onset of miscibility in the bosonic mixture \cite{Ilya}, in the
present case the immiscible components do not form a domain wall, separately
filling two domains partitioned by the wall. Instead, they form a pair of
anti-locked ($\pi $-phase-shifted) spatially modulated density waves, and
spontaneously break the equality of numbers of atoms in the two components
(the latter feature was never observed in the bosonic mixture). The relation
between the imbalance in the spin population, induced by the linear
coupling, and the spatial density-wave patterns developed by the species is
similar to the LOFF phase in a binary fermion gas with unequal values of the
Fermi radius \cite{LOFF}-\cite{LOFFgas}. As the linear-coupling strength, $%
\kappa $, approaches the induced-miscibility threshold, $\kappa _{\mathrm{cr}%
}$, the spatial modulation in the density waves vanishes, and the numbers of
atoms in the components become equal. We have also considered the influence
of inherent asymmetry between the components (the chemical-potential\
difference between them), and dynamical situations, when $\kappa $ is
suddenly switched on (in the symmetric system) from zero to a value
exceeding $\kappa _{\mathrm{cr}}$. In the latter case, the system features
oscillatory relaxation to the mixed state.

\acknowledgments


The work of B.A.M. was supported, in a part, by the Israel Science
Foundation through the grant No. 8006/03. The work of S.K.A. was supported
in part by the FAPESP and CNPq of Brazil.

\end{document}